# Photonic Integrated Neuro-Synaptic Core for Convolutional Spiking Neural Network

Shuiying Xiang[1*], Yuechun Shi[2*], Yahui Zhang[1], Xingxing Guo[1], Ling Zheng[3], Yanan Han[1], Yuna Zhang[1], Ziwei Song[1], Dianzhuang Zheng[1], Tao Zhang[1], Hailing Wang[4], Xiaojun Zhu[5], Xiangfei Chen[6], Min Qiu[7], Yichen Shen[8], Wanhua Zheng[4] & Yue Hao[1]

**Neuromorphic photonic computing has emerged as a competitive computing paradigm to overcome the bottlenecks of the von-Neumann architecture. Linear weighting and nonlinear spiking activation are two fundamental functions of a photonic spiking neural network (PSNN). However, they are separately implemented with different photonic materials and devices, hindering the large-scale integration of PSNN. Here, we propose, fabricate and experimentally demonstrate a photonic neuro-synaptic chip enabling the simultaneous implementation of linear weighting and nonlinear spiking activation based on a distributed feedback (DFB) laser with a saturable absorber (DFB-SA). A prototypical system is experimentally constructed to demonstrate the parallel weighted function and nonlinear spike activation. Furthermore, a four-channel DFB-SA array is fabricated for realizing matrix convolution of a spiking convolutional neural network, achieving a recognition accuracy of 87% for the MNIST dataset. The fabricated neuro-synaptic chip offers a fundamental building block to construct the large-scale integrated PSNN chip.**

[1] State Key Laboratory of Integrated Service Networks, State Key Discipline Laboratory of Wide Bandgap Semiconductor Technology, Xidian University, Xi'an 710071, China. [2] Yongjiang laboratory, No. 1792 Cihai South Road Ningbo, 315202, China. [3]The School of Communications and Information Engineering, Xi'an University of Posts and Telecommunications, Xi'an 710121, China. [4]Laboratory of Solid-State Optoelectronics Information Technology, Institute of Semiconductors, Chinese Academy of Sciences, Beijing 100083, China. [5]School of Information Science and Technology, Nantong University, Nantong, Jiangsu, 226019, China. [6]The College of Engineering and Applied Sciences, Nanjing University, Nanjing 210023, China. [7]Key Laboratory of 3D Micro/Nano Fabrication and Characterization of Zhejiang Province, School of Engineering, Westlake University, Hangzhou 310024, China. [8]Lightelligence, Hangzhou 311121, China. Correspondence and requests for materials should be addressed to S.Y. X. and Y. C. S. (email: syxiang@xidian.edu.cn; yuechun-shi@ylab.ac.cn )

The deep neural network has developed rapidly and achieved record-breaking performance in a broad range of applications, including computer vision, natural language processing, and other fields. These applications produce huge amounts of data that need to be processed, which calls for advanced processors with high speed, high throughput and low latency. While Moore's Law approaches saturation and conventional digital computers face the von-Neumann bottleneck due to the separation of memory and processor units. Neuromorphic computing, which emulates the structure and mechanism of the brain, has emerged as a competitive computation paradigm in the post-Moore era [1-3]. Although the electronic neuromorphic computing systems based on the well-known complementary-metal-oxide-semiconductor (CMOS) technique or emerging electronic devices have made notable progress, they still face limitations in processing speed and energy efficiency [4-11]. As a promising alternative, photonic neuromorphic computing has garnered significant attention as it offers inherent advantages such as ultra-high speed, large bandwidth, and massive parallelism. It has become a prominent research focus in neural computing due to its potential for addressing the limitations of electronic counterparts [12-17].

In addition to the advancements in neuromorphic hardware, novel neural network models have also attracted significant attention. In contrast to traditional non-spiking artificial neural networks (ANNs), spiking neural networks (SNNs) have been proposed and designed to enable low power consumption computing on neuromorphic hardware by using discrete spike signals [18-20]. The implementation of SNNs on a photonic neuromorphic hardware holds great promise for a wide range of applications in latency-critical and power-constrained scenarios, such as autonomous driving and edge computing [21-25].

Spiking neurons serve as the fundamental units of SNNs, and are connected through plastic synapses. In an SNN, the spiking neuron performs the nonlinear spike activation, while the synapse performs the linear weighting function. For the majority of reported photonic neural networks, only linear operation was realized in the photonics domain [26-35]. For instance, two mainstream approaches, namely the coherent synaptic network based on the Mach-Zehnder interferometer (MZI) [27, 29, 32-33] and the incoherent synaptic work based on the microring resonator (MRR) [26, 35] have been widely explored due to their compatibility with CMOS-compatible silicon photonics platform. However, these approaches suffer from serious loss issues and are not well-suited for implementing nonlinear computations directly in the photonic domain. In the reported photonic neural network architectures, nonlinear computations are mainly implemented electronically, relying on high-speed photodetectors (PDs) and analog-to-digital (AD) converters to convert the optical linear computation results back to the digital domain. Such hybrid architectures, consisting of photonic synapse and electronic spiking neurons, present obstacles for the photonic implementation of multi-layer or deep neural networks. The frequent optic-electro (OE) and electro-optic (EO) conversions, as well as the AD and digital-to-analog (DA) conversions, hinder the seamless integration of photonic functionalities across multiple layers.

Tremendous efforts have been made to realize photonic spiking neurons, aiming to enable all-optical photonic SNNs without the need for frequent OE/OE and AD/DA conversions. Among various simplified spiking neuron models in computational neuroscience, the leaky integrate-and-fire (LIF) model has gained popularity due to its simplicity. Optical implementations of the LIF neuron have received considerable attention [36]. Several optical implementations of the LIF neuron have been explored, including polarization switching vertical-cavity surface-emitting lasers (VCSELs) [37-39], VCSEL with a saturable absorber (VCSEL-SA) [40-41], graphene excitable laser [42], hybrid integrated phase-change material (PCM) and MRR [21, 43], micropillar lasers [44], integrated distributed feedback (DFB) laser and PD [45], passive microresonator [46], and integrated Fabry–Perot laser with a saturable absorber (FP-SA) [25, 47]. However, these photonic spiking neurons have been designed and fabricated independently from the photonic synapses, limiting the scalability of fully-functional photonic SNNs.

Here, we proposed, fabricated and demonstrated a photonic neuro-synaptic chip that enables simultaneous implementation of linear weighting and nonlinear spiking activation based on a photonic integrated DFB laser with an intracavity saturable absorber (DFB-SA) for the first time. By precisely tuning the external optical injection and the time-varying bias current of the gain region of the DFB-SA, we experimentally demonstrate both the intrinsic excitability plasticity and synaptic plasticity in the same DFB-SA. The nonlinear spike activation and linear weighting can be simultaneously realized in a single DFB-SA chip. In addition, parallel linear weighting and nonlinear spike activation were demonstrated in a constructed prototypical neuromorphic photonics system consisting of two pre-synaptic DFB-SAs and one post-synaptic FP-SA. To demonstrate the scalability of the DFB-SA, a four-channel photonic neuro-synaptic array was further fabricated and applied to perform dot product between two vectors. Through time-multiplexing matrix convolution, we successfully demonstrated pattern classification tasks. We further developed a theoretical model to numerically simulate the photonic integrated DFB-SA, validating our experimental findings. The integration of linear weighting and nonlinear spike activation in a single chip opens up new possibilities for efficient and scalable photonic neuromorphic computing, presents a promising building block for constructing multi-layer photonic SNN hardware within the InP integration platform.

## Results
### Principle for photonic neuro-synaptic core.
In a biological neural network, the neurons are connected by plastic synapses. As depicted in Fig.1 (a), the dendrites receive external stimuli or pre-synaptic spikes, while the soma converts these inputs into spike events and performs nonlinear spiking activation. The synapses adjust the connection strength, commonly referred to as weight in ANNs. The weighted spike is then transmitted to the dendrites of the neighboring neurons. To address the challenge of implementing both linear weighting and nonlinear spiking activation in a single device, we proposed, designed and fabricated a photonic integrated DFB-SA, which we refer to as a photonic neuro-synaptic chip. The structure of the DFB-SA chip is illustrated in Fig.1 (b). The grating is designed with a sampled grating. We shift the half period of the sampling structure in the middle of the gain region of the DFB laser, which can equivalently introduce a π phase shift (π-EPS). This grating structure can be fabricated with the reconstruction-equivalent-chirp (REC) technique, which allows for a large-scale DFB array with high wavelength precision [48]. Additionally, anti-reflection (AR) and high-reflection (HR) coating are applied to the two laser facets to enhance the light emission power, and the SA region is positioned near the HR side. Figure 1(c) provides an overview of the spike processing principle in the DFB-SA. The gain region of the DFB-SA is driven by a current source, denoted as the gain current $I_G$, while the SA region is reversely driven by a voltage source, denoted as $V_{SA}$. By applying a time-varying current to the gain region, we achieve dynamically linear weighting functionality. The interaction between photons and electrons in the gain and SA regions enables the emulation of a LIF-type spiking neuron.

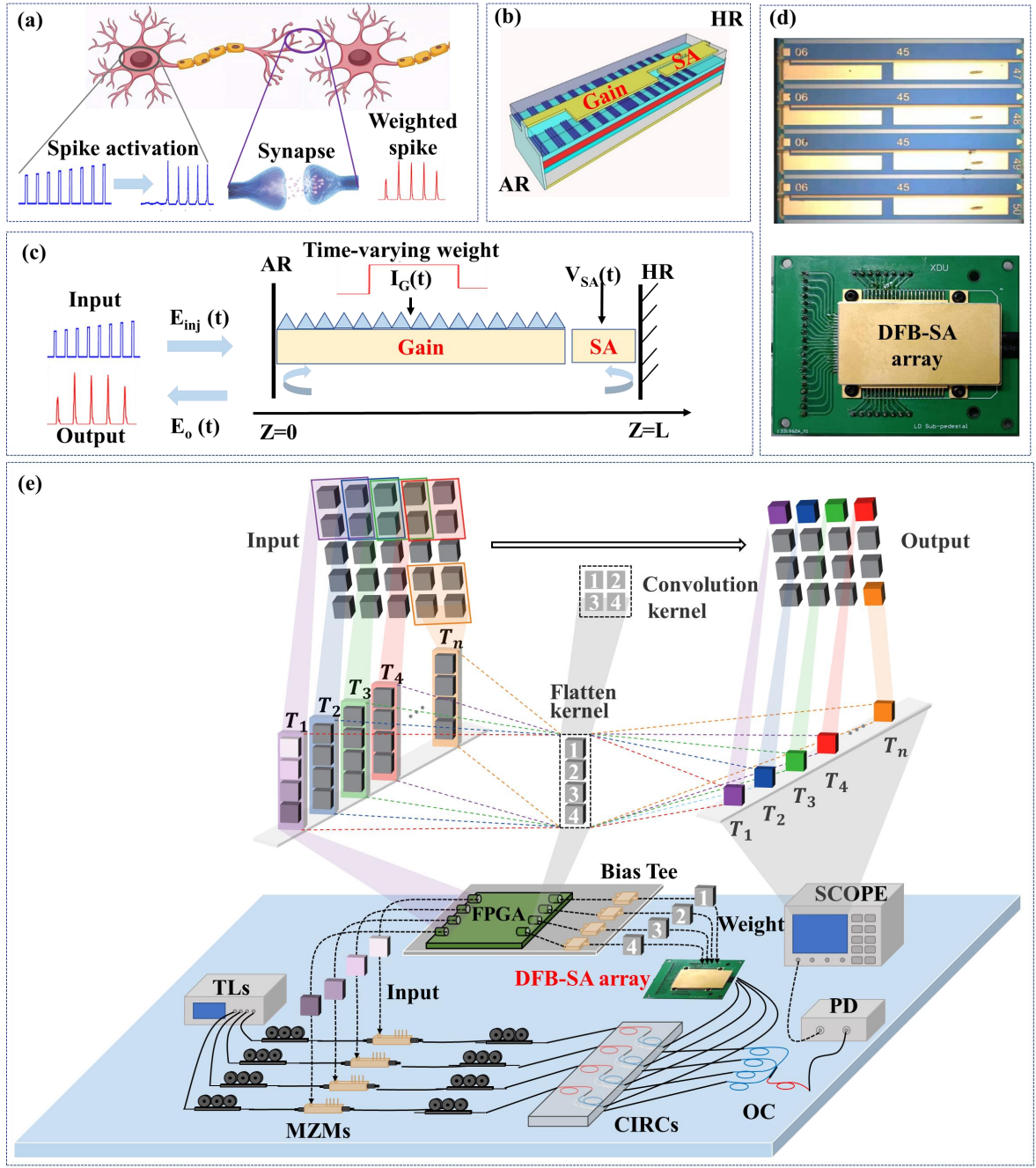

Fig. 1. The operation principle of photonic neuron-synaptic core. (a) The schematic diagram of biological neuron, (b) the structure and micrograph picture of the designed DFB-SA chip, (c) the spike processing in a DFB-SA, (d) the micrograph picture of the fabricated DFB-SA array chip and the compact packaged module, (e) spike-based matrix convolution based on the four-channel DFB-SA array.

We further designed and fabricated a four-channel photonic neuro-synaptic on-chip array to implement two-dimensional (2D) convolution for spiking CNNs. To fully leverage the temporal dynamics of the SNN, we have demonstrated time-multiplexing matrix convolution. The fabricated four-channel DFB-SA array and the compact packaged module are depicted in Fig.1 (d). The principle of using the DFB-SA array to achieve spike-based matrix convolution is illustrated in Fig.1 (e). The input and kernel matrices are generated by the field-programmable gate array (FPGA). The input is electro-optically modulated and optically injected into the DFB-SA array, while the weight is directly imported to the bias current of the gain region. Thus, the on-chip DFB-SA array performs dot product operations and acts as a photonic dot product core. For detailed information on the experimental setup, please refer to the Method section.

**Intrinsic neuron excitability plasticity**.

The experimental setup for emulating the neuronal intrinsic plasticity and linear weighting using a DFB-SA is presented in Fig. 2 (a). For detailed information regarding the setup, please refer to the Method section. In our experiment, we maintained a fixed temperature of 25°C. The threshold current of the DFB-SA was measured to be approximately $I_G$=86 mA for $V_{SA}$=0 V and $I_G$=94 mA for $V_{SA}$= − 0.4 V. The optical spectrum of a free-running DFB-SA is presented in Fig. 2(b). The side mode suppression ratio was found to be approximately 50.9 dB, indicating a high level of suppression of unwanted side modes.

For a LIF spiking neuron, temporal integration, spike threshold and generation as well as the refractory period are crucial information processing mechanisms. Here, we successfully demonstrated these spike-based nonlinear processing mechanisms using the fabricated DFB-SA chip. As shown in Fig. 2(c), we designed five input pulses with varying amplitudes for a given external stimulus strength and injection power, and obtained different nonlinear spike activations simply by adjusting the gain current. For instance, when $I_G$=88 mA, only the stimulus pulse with the highest amplitude exceeded the excitability threshold, resulting in a successful response spike. With the increase of gain current, the number of responded spikes increased in a completely controllable manner. Note that, the external stimulus strength corresponding to the synapse weight remained fixed throughout the experiment. The different spike response was solely achieved by adjusting the threshold of the DFB-SA spiking neuron. This demonstrates the presence of intrinsic excitability threshold plasticity in the fabricated DFB-SA, which is similar to its biological counterpart [49]. Thus, nonlinear spike activation for a photonic SNN can be achieved in the DFB-SA.

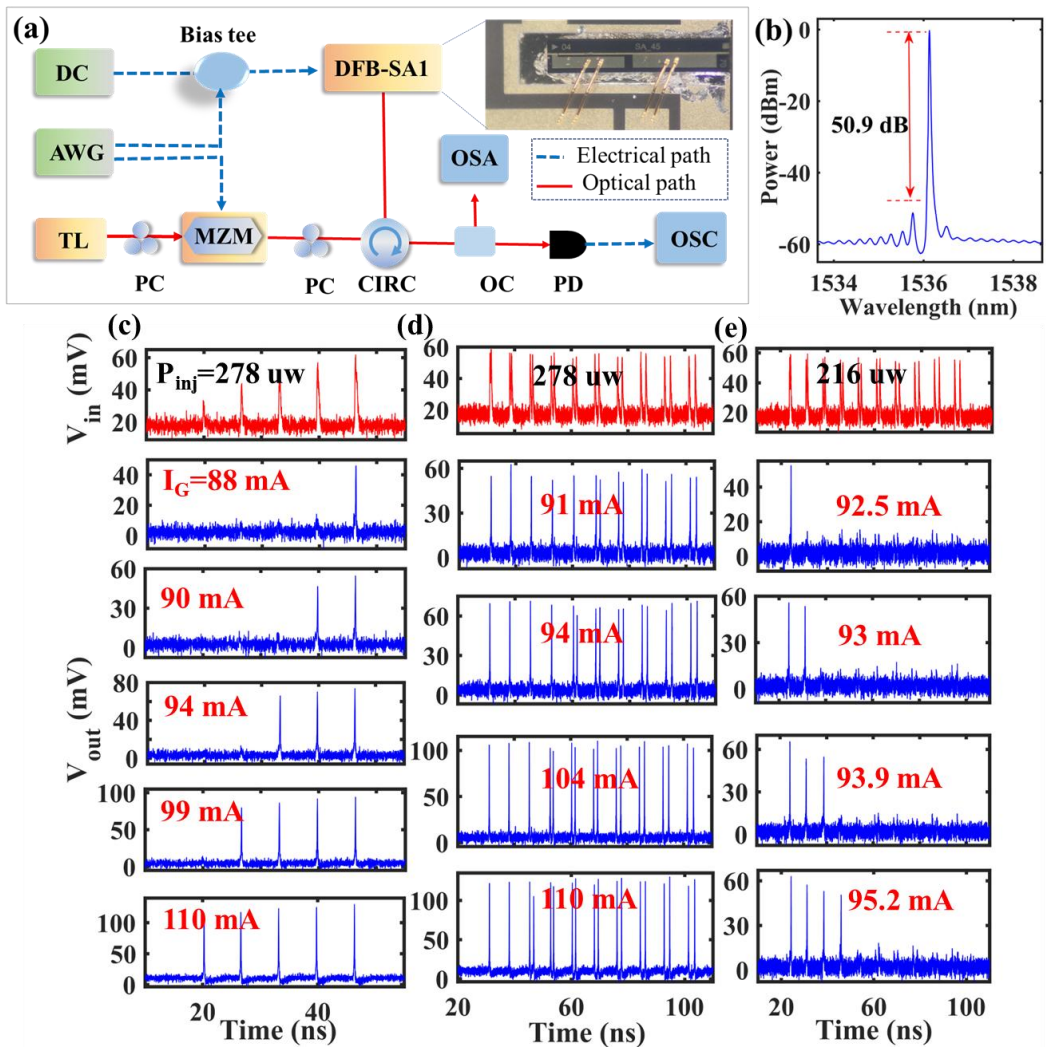

Fig. 2. Experimental demonstration of intrinsic excitability plasticity in a single fabricated DFBSA chip. (a) Experimental setup of a DFB-SA for emulating the neuronal intrinsic plasticity and linear weighting, (b) optical spectrum of the free-running DFB-SA, (c) different excitability threshold, (d) different refractory period, and (e) different temporal integration behavior under different gain currents. $V_{SA}$=−0.4V. The inset in (a) corresponds to the wire bonding of the fabricated DFB-SA chip.

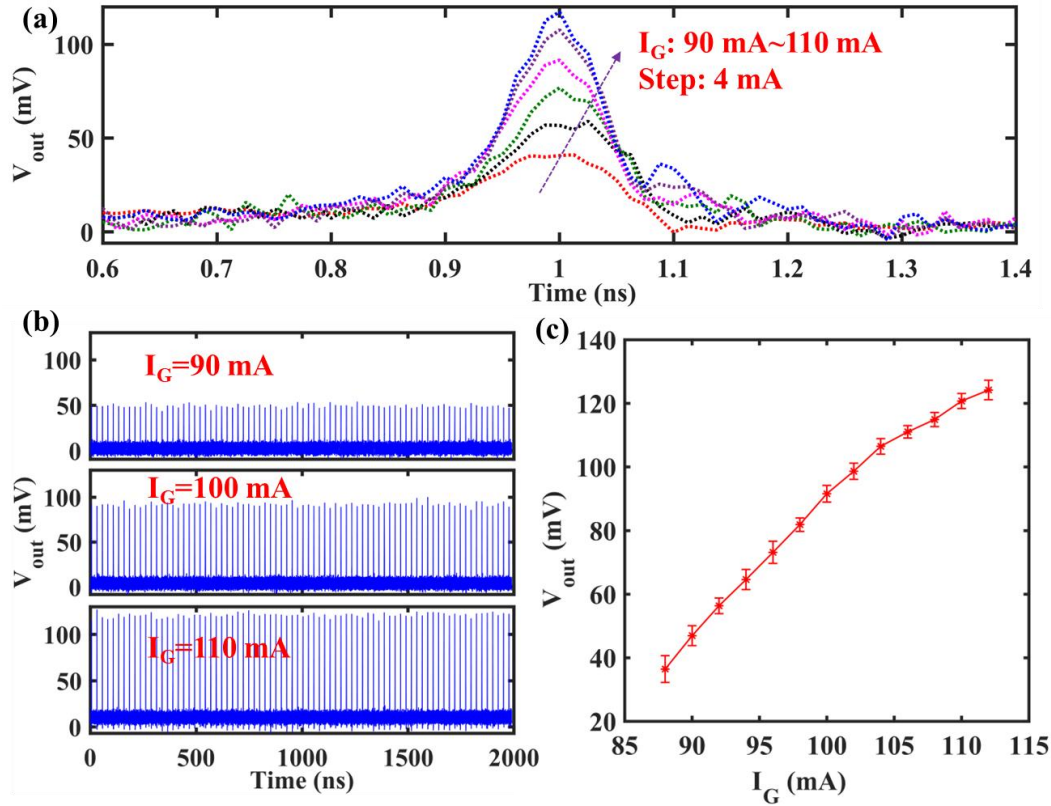

Fig. 3. Continuously tunable weighted spike output for different static gain currents in a single fabricated DFB-SA chip. (a) The weighted output of a single spike for different gain currents, (b) the weighted spike trains for three representative gain current, (c) the spike peak amplitude as a function of gain current.

Similarly, we fixed the external stimulus strength, and designed 10 pulse pairs with increasing inter-spike interval (ISI) to explore the intrinsic plasticity of the refractory period and temporal integration behavior. As presented in Fig. 2(d), for different input pulse pairs, the number of response spikes was either 1 or 2 depending on the ISI. For $I_G$=91 mA, the first 5 pulse pairs with relatively small ISIs resulted in a single response spike, while the last 5 pulse pairs with larger ISIs triggered two response spikes each. With the increase of $I_G$, the number of pulse pairs leading to a single response spike gradually decreased to 4, 3 and 2, respectively. To demonstrate the temporal integration behavior, the injecting power was decreased to a level where a single stimulus pulse alone could not elicit a response spike. In Fig. 2(e), when $I_G$=92.5 mA, only the first pulse pair with the smallest ISI could be temporally integrated, exceeding the excitability threshold and eliciting a response spike. With the increase of gain current, the number of response spikes also increased. In other words, the response spike resulting from the temporal integration effect could be controlled by simply adjusting the gain current of the DFB-SA.

Note, the demonstrated intrinsic neuron excitability plasticity implies that, when developing a supervised training algorithm for a photonic SNN consisting of the DFB-SA, not only the weights can be considered as adjustable parameters, but the threshold of the DFB-SA can also be trainable to accelerate the training process. Thus, a novel hardware-aware training algorithm that combines both the weight and threshold modulation is highly desirable to enhance the performance of photonic SNNs.

**Continuously tunable synaptic plasticity.**

Next, we experimentally demonstrated the linear weighting function of the fabricated DFB-SA chip by employing a statically-varying bias current. We defined a periodic optical pulse train as the external stimulus and adjusted the gain current, ensuring that the DFB-SA operated in an LIF-like response manner. As illustrated in Fig. 3 (a), the peak amplitude of the response spike gradually increased as the gain current was increased, ranging from 90 mA to 110 mA in 4 mA increments. In Fig. 3(b), we present long time traces of the weighted optical spike trains for $I_G$=90 mA, 100 mA and 110 mA. It is verified that a higher gain current contributes to a larger spike amplitude. Moreover, for a given gain current, the amplitude of the weighted spike remained almost constant across different periods, indicating the stability of the weighting function. We further present the peak amplitude of the response spike as a function of the gain current in Fig. 3(c). The bias current varied from 88 mA to 112 mA. Notably, the peak amplitude of the weighted spike increases almost linearly with the gain current, which can be attributed to the gain-switched operation mechanism. Consequently, in a photonic SNN, the synapse weight can be readily mapped to the gain current of the DFB-SA, facilitating straightforward weight modulation.

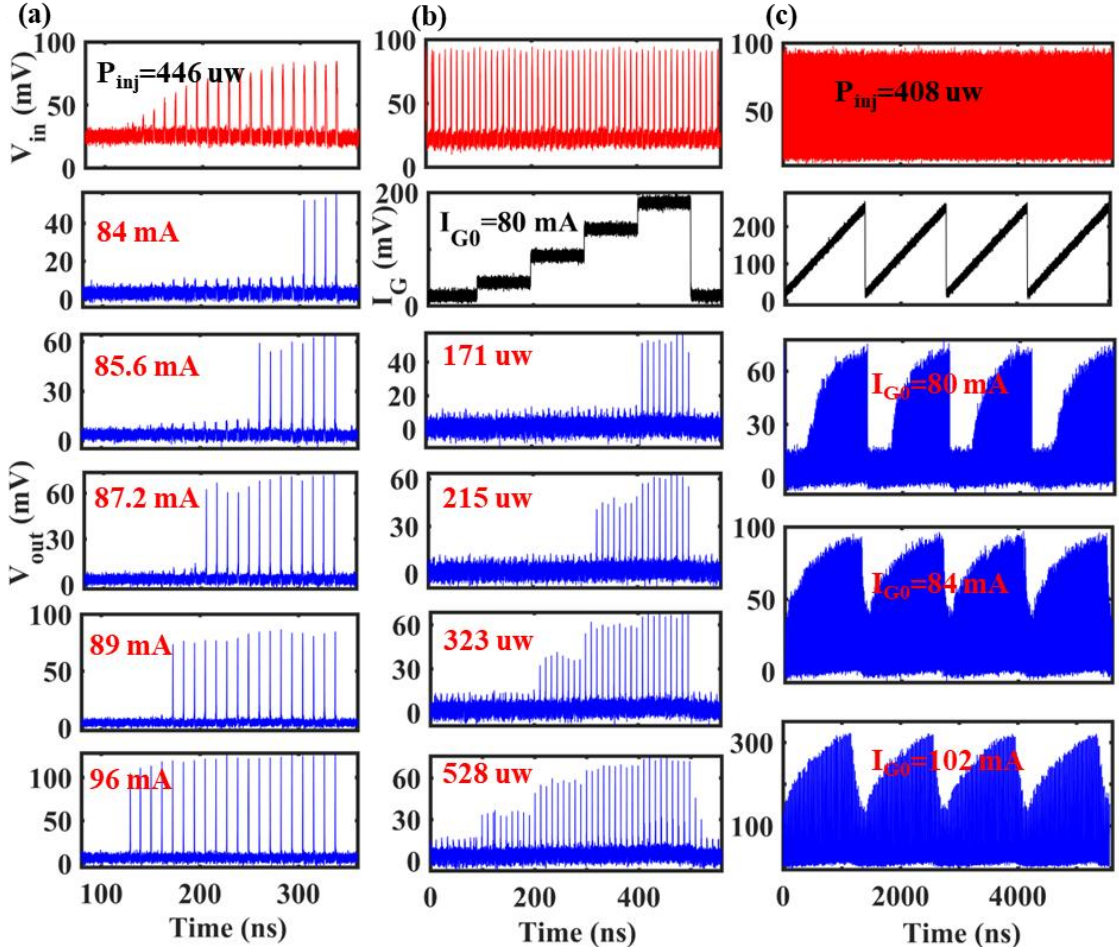

Fig. 4. Experimental demonstration of simultaneous implementation of neuron excitability threshold and synaptic linear weighting functions in a single DFB-SA. In (b) $I_{G0}$=80 mA, $I_G = I_M + I_{G0}$. (a) corresponds to static gain current at different level, (b) and (c) correspond to step-wise and continuously-varying gain current, respectively.

By simply tuning the bias current of the gain region of the DFB-SA, we successfully achieved precise control over the amplitude of the weighted optical spikes, thus enabling a photonic integrated spiking neuron with incorporated spike weighting capability. Note, in comparison to biological counterparts that operate on millisecond timescales, the photonic neuro-synaptic unit utilizing the fabricated DFB-SA offers significantly faster operation speeds in the sub-nanosecond range, thanks to the short carrier lifetimes inherent to these chips. Therefore, the proposed photonic neuron-synaptic unit avoids the use of additional photonic weighting elements, making it highly desirable for the further scalability of photonic SNN hardware.

**Simultaneous implementation of nonlinear spike activation and linear weighting.**

In the following, we considered the scenario where a dynamically time-varying gain current was applied to the DFB-SA, allowing for the simultaneous implementation of nonlinear spike activation and linear weighting. Note, the nonlinear spike activation is implemented due to the excitability threshold modulation. For the purpose of comparison, we employed a time-varying optical input stimulus and considered different constant gain currents, as depicted in Fig. 4 (a). It is evident that, for a given gain current, as long as the stimulus intensity exceeds the excitability threshold, the spike amplitude remained nearly constant regardless of varying stimulus strengths. In addition, the excitability threshold was decreased with the increase of gain current, enabling a greater number of input pulses that can trigger response spikes. However, it is important to note that the spike amplitude differs across different gain currents, with larger gain currents resulting in higher spike amplitudes.

Then, we considered a fixed input strength and applied a dynamically modulated gain current. In this experiment, the repetition rate of the input pulse train was set as 0.5 GHz. As presented in Fig. 4(b), the gain current was modulated using 5 discrete constant levels with an arbitrary waveform generator (AWG) corresponding to 5 distinct weight values. The modulated current was then combined with a static bias current via a bias tee. The total gain current can be expressed as $I_G = I_{G0} + I_M$, where $I_{G0}$ denotes the static bias current and $I_M$ represents the modulated bias current. It is evident that each case exhibited a different excitability threshold. When the injection power was 171 μw, only the highest level of $I_M$ resulted in spike output. Increasing the injection power to 215 μW led to weighted spikes with two distinguishable amplitude levels. Further increasing the injection power enabled the attainment of three or four clusters of weighted spikes, each with varying spike amplitudes. In addition, we also considered the scenario of continuously time-varying gain current. As depicted in Fig. 4(c), the excitability threshold plasticity and linear weighting function was simultaneously achieved by simply modulating the gain current of the DFB-SA. Thus, the DFB-SA enables nonlinear spike activation and high-speed (0.5 GHz) dynamic weight tunability. The enlargement of a single period in Fig. 4(c) is further illustrated in the Supplement material Note 1.

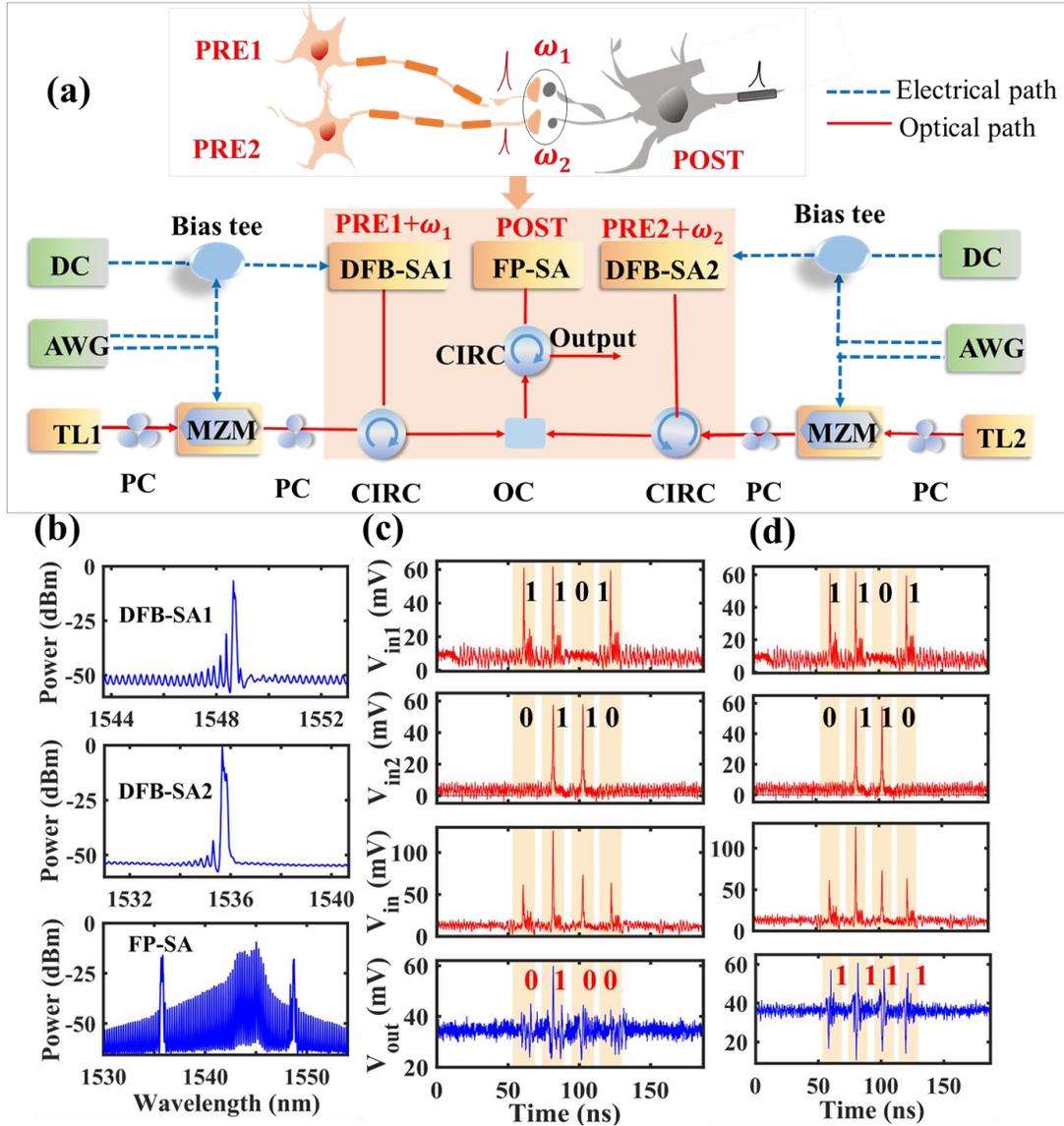

Fig. 5. Spike-based neural computation in a photonic SNN consisting of two PRE DFBSAs and one POST FPSA. (a) Experimental setup of a full-functional prototypical photonic neuromorphic system, (b) optical spectra of DFB-SAs and FP-SA, (c) spike-based AND, (d) spike-based OR.

We also considered other input spike trains with higher repetition rates, and found that the linear weighting function was degraded. Some spikes cannot be normally weighted, which is limited by the refractory period. Note, the refractory period is mainly determined by the carrier lifetime, which can be reduced by using a shorter cavity length.

We further developed a theoretical model based on the well-known time-dependent traveling wave model to numerically reproduce the linear weighting function and nonlinear spike activation [50]. The model and its corresponding parameters are detailed in the Supplement material Note 2. The numerical results obtained from the model are presented in the Supplement material Note 3. These results demonstrate that different gain currents result in distinct excitability thresholds. In addition, the modulation of gain current produces weighted spikes with distinguishable amplitude levels. Namely, the numerical results agree well with the experimental findings, thus providing a computational model that facilitates hardware-algorithm co-design and optimization of photonic SNNs utilizing the DFB-SAs.

**Fully-functional photonic neuromorphic prototypical system.**

We further constructed a fully-functional prototypical photonic neuromorphic system utilizing the fabricated DFB-SAs for

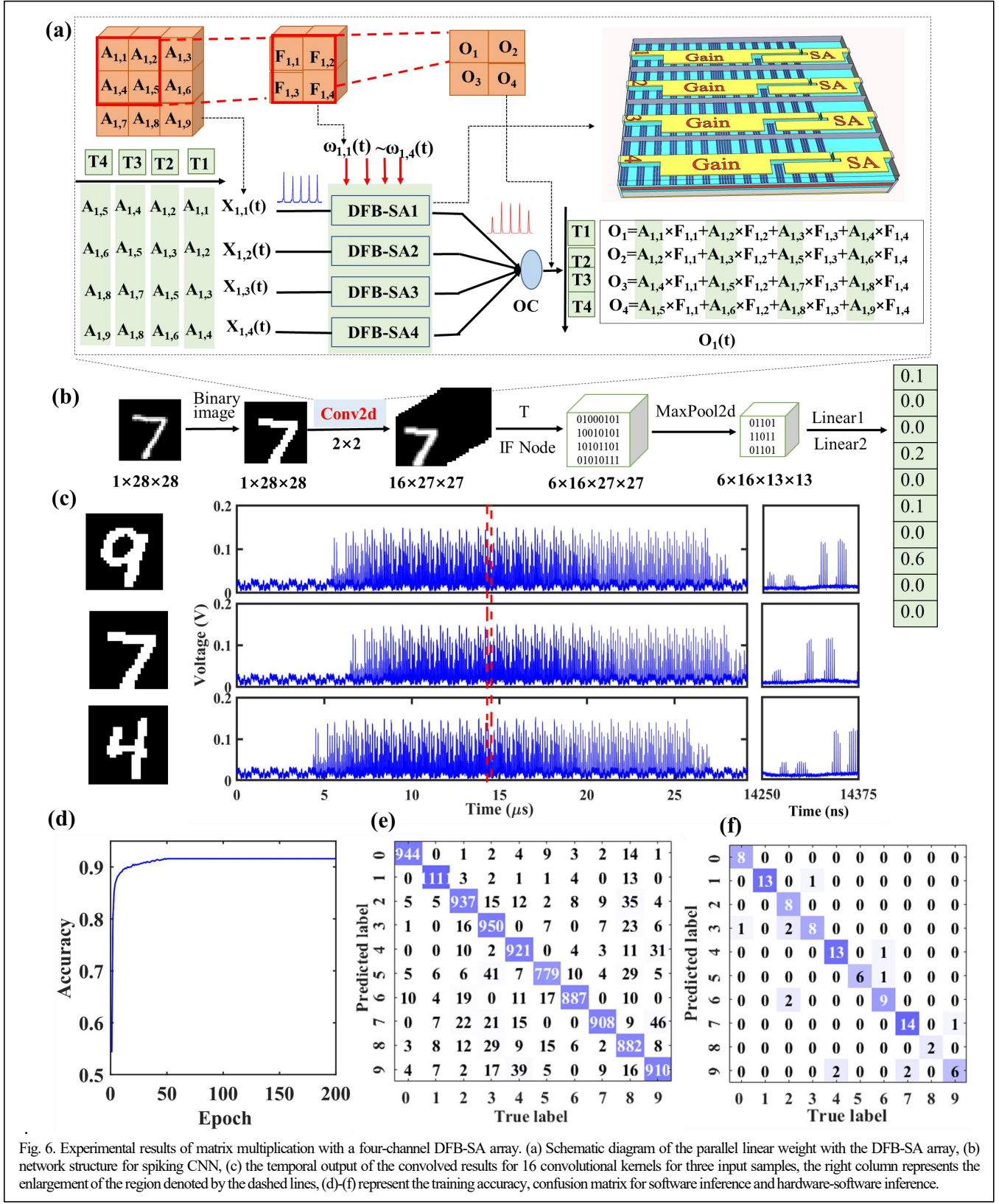

Fig. 6. Experimental results of matrix multiplication with a four-channel DFB-SA array. (a) Schematic diagram of the parallel linear weight with the DFB-SA array, (b) network structure for spiking CNN, (c) the temporal output of the convolved results for 16 convolutional kernels for three input samples, the right column represents the enlargement of the region denoted by the dashed lines, (d)-(f) represent the training accuracy, confusion matrix for software inference and hardware-software inference.

parallel linear weighting and nonlinear spike activation. Note, as the packaged DFB-SAs operate at different wavelengths, we adopted an FP-SA as the post-synaptic neuron because it can support wide bandwidth operation [25]. The system configuration is illustrated in Fig. 5(a), where DFB-SA1 and DFB-SA2 act as pre-synaptic photonic spiking neurons and the corresponding weighting devices, while the FP-SA serves as the post-synaptic photonic spiking neuron. The weighted spike outputs of the DFB-

SAs were optically combined and injected into the FP-SA. The optical spectra of the DFB-SAs and FP-SA are displayed in Fig. 5 (b). In this setup, the binary patterns 1101 and 0110 are employed as inputs for the DFB-SA1 and DFB-SA2, respectively. By setting different gain currents of the DFB-SAs, they performed digital-to-spike conversion and parallel linear weighting. The optical coupler then performed the summation of the two weighted spikes. The weighted sum was subsequently injected into the FP-SA for nonlinear spike activation. The output of the FP-SA reveals that it can yield either 0100 or 1111, depending on the weighting conditions of the DFB-SAs or the excitability threshold of the FP-SA. Consequently, the constructed neuromorphic prototypical system can be flexibly reconfigured to execute spike-based AND as well as OR operations. Note, both the weighted addition operation and nonlinear spike activation are achieved within this photonic neuromorphic prototypical system based on the DFB-SAs and FP-SA, which can be integrated onto the same InP integration platform.

**Time multiplexing matrix convolution with four-channel DFB-SA array.**

To demonstrate the scalability of the proposed photonic neuron-synaptic unit based on the DFB-SA, we fabricated a four-channel DFB-SA array and applied it to perform matrix convolution for a convolutional SNN. The operation principle is illustrated in Fig. 6 (a). In this configuration, the convolutional kernel size is 2×2, the stride step is 1. Each element in the convolutional kernel is mapped to a respective DFB-SA within the array. The gain currents of the fabricated DFB-SAs array are configured according to the trained kernels. The convolved results are obtained as the output of the optical coupler. The layout of the DFB-SA array is shown in the inset of Fig.6 (a), which exemplifies the potential for scalability of the photonic neuron-synaptic unit.

The spiking CNN was trained using a digital computer to classify the MNIST dataset, which consists of handwritten digits ranging from 0 to 9. The training set contains 60,000 samples and the test set contains 10,000 samples. The size of each image is 28×28 pixels. The spiking CNN consists of a convolutional layer with 16 convolutional kernels, an activation layer, a pooling layer, and two fully-connected layers, as depicted in Fig. 6(b). The trained 16 convolutional kernels were presented in Supplemental material Note 4. In this experiment, a hardware-software collaborative approach was adopted for inference. The photonic hardware was utilized to realize the convolutional layer. For evaluation, 100 test images were considered, and the experimental results for some representative input samples are presented in Fig. 6(c) (Please refer to Supplemental material Note 5 for additional input samples). The 16 convolutional kernels were time-multiplexed to take full advantage of the dynamically weight modulation capability of the DFB-SA array. The sampling rate of the DA was set at 6.4 Gbits/s. Considering the refractory period, 15 zeros were inserted between two pixels to match the FPGA bitwidth, resulting in a temporal duration of each pixel of 1/6.4Gbits/s×16=2.5 ns. For each image, the temporal output of the convolved results for the 16 convolutional kernels includes 27×27×16 pixels. Thus, the total length of each measured output time series is 2.5 ns×27×27×16=29.16 μs. To demonstrate fast dynamically weight modulation, we calculated the convolved results row by row. Each row of pixels was convolved with 16 different weights corresponding to different convolutional kernels. As a result, the temporal duration of a fixed weight was 2.5 ns×27=67.5 ns, yielding a modulation rate of approximately 14.8 MHz. Post-processing was performed with a digital computer. The feature maps are displayed in Supplemental material Note 6. After subsequent processing steps, the results indicate that pattern classification with a recognition accuracy of 87% was achieved. Note, with the same network structure, by using pure software inference, the recognition accuracy is 92.29% (for 10000 test images) and 90% (for 100 test images). The training accuracy, confusion matrix for software inference and hardware-software inference can be found in Figs.6 (d)-(f). The 3% loss in inference accuracy observed in the hardware-software inference scenario may be attributed to noise present in the experimental setup.

We further conducted numerical simulations to explore time-multiplexing parallel dot product using our developed time-dependent traveling wave model. The numerical results of four-channel parallel linear weighting are presented in Supplemental material Note 7. The simulations revealed that the linear weighting operation could be effectively realized in parallel using this configuration. Moreover, we considered a larger convolution kernel size and simulated the case of a 3×3 convolution kernel. The numerical results of 9-channel parallel linear weighting are provided in Supplemental material Note 8. It indicates that, 9-channel parallel dot product could be achieved by employing an array of nine DFB-SAs as the photonic dot product core. We further consider a spiking CNN network similar to the one shown in Fig. 6(b) but with 3×3 convolution kernels. The simulation results are presented in Supplemental material Note 9. The inference accuracy is 92.45% for the spiking CNN comprises one convolutional layer. Additionally, spiking CNN networks with two convolutional layers are also simulated, and the inference accuracy is 93.76% and 94.42% for convolution kernel size of 2×2 and 3×3, respectively. The numerical findings validate the feasibility and scalability of the proposed photonic neuron-synaptic unit for larger convolutional operations.

**Discussions**

The computation speed of the linear weighting operation in the fabricated four-channel DFB-SA array, with statically configured weights, can be estimated at 2×4×10 G=80 GOPS/s. On the other hand, the nonlinear computation speed for a single channel is approximately 2 G Spike/s, limited by the refractory period. The energy consumption per spike is approximately 19.99 pJ. The area occupied by a single DFB-SA chip is around 1500 μm ×300 μm=0.45 mm$^2$. It is important to note that there is significant potential for improving these metrics. Further optimization can be achieved by reducing the threshold of the DFB-SA to around or below 5 mA, and by reducing the area of a single DFB-SA chip to 300 μm×127 μm. The reconfigurable rate, which is associated with the speed of dynamics weight update, is estimated at 0.5 GHz, significantly faster than that of the thermos-optic phase shifters based on silicon photonics. Moreover, the dynamical weighting speed can be further increased to approximately 10 GHz by reducing the cavity length of the DFB-SA to 300 μm, which is compatible with the time-multiplexing matrix convolution requiring fast weight modulation. Additionally, a previously demonstrated 60-channel DFB array with high wavelength precision [48] can substantially increase the computation speed due to its high parallelism. In such a configuration, a computation speed of 2×60×10 G = 1.2 TOPS/s can be expected.

Compared to silicon photonics-based weighting elements, the proposed DFB-SA offers several advantages. Firstly, it eliminates the loss issue as gain amplification can be easily achieved in the InP integration platform. This characteristic is particularly beneficial for the implementation of multi-layer photonic SNNs. On the other hand, weight control using thermo-optic phase shifters in the silicon photonics platform is limited to a tuning rate in the kHz range. In contrast to the weighting devices based on the VCSEL biased below the lasing threshold [51-52], the DFB-SA is biased above the lasing threshold but below the self-pulsating threshold. This allows for efficient spike amplification using the gain provided by the DFB-SA, which is crucial for cascaded propagation of weighted spikes. Moreover, the DFB-SA enables the simultaneous realization of both linear weighting and activation functions in a single device. This capability avoids frequent OE/EO and AD/DA conversions among different layers in a photonic weighting and electronic activation architecture, and alleviates the inherent challenge of optical coupling that arises when attempting to implement large-scale all-optical SNNs through hybrid integration of weighting and activation elements fabricated using different materials and devices [53-54].

In conclusion, we have proposed and successfully fabricated a novel photonic spiking neuron chip based on an integrated DFB-SA, which enables simultaneous spike activation and linear weighting functions for the first time. This chip has a simple structure and can be readily integrated on a large scale using commercially mature fabrication processes available in photonics foundries. By adjusting the gain current applied to the DFB-SA, the chip can be flexibly reconfigured to function as a linear weighting device or nonlinear spike activation device. The fully-functional neuromorphic prototypical system, comprising DFB-SAs and FP-SA, successfully performed spike-based AND as well as OR operations. Furthermore, using a fabricated four-channel DFB-SA array, we benchmarked the hardware-software collaborative inference on the MNIST dataset, achieving an inference accuracy of 87%. Overall, our work demonstrates the potential of the integrated DFB-SA for advancing the field of photonic neuromorphic computing, offering scalability of fully-functional integration of lossless multilayer or deep photonic SNN in a single integrated chip.

## Methods
### Experimental setup.

The experimental setup for testing the four-channel DFB-SA array is illustrated in Fig. 1 (e). The input and weight signals were generated using an FPGA (Zynq UltraScale+ RFSoC ZCU216) equipped with a high-speed AD/DA array. The FPGA was controlled by a digital computer. To generate optical carriers, four-channel continuous-wave tunable laser sources were utilized. The input signals were modulated using the Mach-Zehnder modulator (MZM). Polarization controllers were employed to align the polarization state. The modulated outputs of the MZMs were then optically injected into the DFB-SA array via three-port optical circulators. The kernel signals were applied to the gain current of the DFB-SA arrays via bias tees. The outputs of four DFB-SAs were combined by a four-port optical coupler. The combined signal was detected by a photodetector (PD, Agilent/HP 11982A) and then recorded by an oscilloscope (OSC, Keysight DSOV334A).

The experimental setup for testing a single DFB-SA is presented in Fig. 2 (a). An AWG (Tektronix AWG70001A) was used to generate the electronic stimulus and the time-varying modulated current for the DFB-SA. The time-varying current was combined with a direct current source using a bias tee and then applied to the gain region of DFB-SA. A tunable laser (TL, AQ2200-136 TLS module) provided the optical carrier. The electro-optic conversion was realized using an MZM. The modulated optical stimulus was then injected into the gain section of the DFB-SA via an optical circulator. The output of the DFB-SA neuron was analyzed by an optical spectrum analyzer (OSA, Advantest Q8384), and meanwhile, it was also detected by the PD and recorded by the oscilloscope.

In addition, the experimental setup for testing the prototypical photonic neuromorphic system, consisting of two DFB-SAs and one FP-SA is illustrated in Fig. 5 (a). The setup for DFB-SA2 is identical to that of DFB-SA1. An optical coupler is employed to combine the weighted spikes generated by the two DFB-SAs, and an optical circulator is utilized to inject the combined weighted spikes into the FP-SA and receive the spike activation output from the FP-SA.

**Model of the DFB-SA.** In order to gain a deeper understanding of the intrinsic excitability plasticity and synaptic plasticity in the DFB-SA, we developed a comprehensive model of the DFB-SA based on the time-dependent coupled-wave equations. We modified the model to incorporate the gain region, SA region, and an external optical injection term [26-27, 50]. The rate equations governing the carrier density in both the gain and absorber regions, and the coupled-mode equations describing the behavior of the forward and backward traveling waves, are detailed in the Supplement material Note 2.

**Data availability**. The data that support the findings of this study are available from the corresponding author upon request.

## Acknowledgements

This work was supported by the National Key Research and Development Program of China (2021YFB2801900, 2021YFB2801901, 2021YFB2801902, 2021YFB2801904); National Natural Science Foundation of China (No. 61974177, No.61674119); National Outstanding Youth Science Fund Project of National Natural Science Foundation of China (62022062); The Fundamental Research Funds for the Central Universities (QTZX23041).

## Author contributions

S. Y. X. and Y. C. S. designed the experiments. Y. C. S., X. F. C., and H. L.W. designed, optimized and fabricated the devices. Y. H. Z., X. X. G., L. Z, D. Z. Z., performed experimental measurements. Y. N. H., Z.W. S. and T. Z. performed the simulations. S. Y. X., Y. C. S., X. J. Z., M .Q., Y. C. S. prepared the manuscript. S. Y. X., W.H. Z, and Y. H. directed all the research. All authors analyzed the results and implications and commented on the manuscript at all stages.

## Competing interests

The authors declare no competing interests.

## Supplementary information

Supplementary information is available for this paper at XXX.